# Elementary analysis of interferometers for wave-particle duality test and the perspective of going beyond the complementarity principle

**Zhi-Yuan Li (李志远)**

Laboratory of Optical Physics, Institute of Physics, Chinese Academy of Sciences, Beijing 100190, China

Email address: lizy@aphy.iphy.ac.cn

**Abstract**

Wave-particle duality and complementarity principle stand at the conceptual core of quantum theory in its orthodox Copenhagen interpretation. They imply that the wave behavior and particle behavior of quantum objects are mutually exclusive to each other in experimental observation. Here we make a systematic analysis using the elementary methodology of quantum mechanics upon Young's two-slit interferometer and Mach-Zehnder two-arm interferometer with the focus placed on how to measure the interference pattern (wave nature) and which-way information (particle nature) of quantum objects. We design several schemes to simultaneously acquire the which-way information for an individual quantum object and the high-contrast interference pattern for an ensemble of these quantum objects by placing two sets of measurement instrument that are well separated in space and whose perturbation on each other is negligibly small within the interferometer at the same time. Yet, improper arrangement and cooperation of these two sets of measurement instrument in the interferometer would lead to failure of simultaneous observation of wave and particle behavior. The internal freedoms of quantum object could be harnessed to probe both the which-way information and interference pattern for the center-of-mass motion. That quantum objects can behave beyond the wave-particle duality and complementarity principle would stimulate new conceptual examination and exploration of quantum theory at a deeper level.

# 1. Introduction

Quantum theory has made brilliant success in predicting the physical and chemical properties of microscopic objects such as electrons, atoms, molecules, and others, and macroscopic objects such as crystals, insulators, semiconductors, superconductors, and others. There is no doubt that quantum theory is one of the most deliberate and successful physical theory in human history. However, the interpretation of quantum theory has been an issue ever since its founding nearly 100 years ago [1-18] and has raised extensive hot controversies, e.g., between Einstein and Bohr [1-3,5]. The wave-particle duality of quantum objects, and more generally, the complementarity principle, stands on the central conceptual core of quantum theory. According to the orthodox Copenhagen interpretation, all quantum objects including massless and massive particles exhibit mutually exclusive behaviors of two intrinsic attributes of the wave nature and particle nature, namely, they behave either as wave or as particles, depending on how they are observed and measured, but never both [1-3,9,10,12-18].

In recent years there are still many theoretical and experimental discussions and investigations on the conceptual implication of the wave-particle duality of various quantum objects, such as photons, electrons, atoms, molecules, and even macromolecules and heavy particles [19-27]. Besides, a wide variety of experimental setup, either gerdanken (thought) or practical, have been adopted [24-42]. All results, without any exception, just confirm once and once again the old wisdom that the wave-particle duality stands solidly and firmly at the conceptual center of quantum theory. In this paper we will show a different perspective, by making an elementary quantum mechanical analysis of previous gedanken or realistic wave-particle duality experiments, that it is possible to design experiments to demonstrate that quantum objects can behave both as wave and as particle. We will focus our analysis on two most popular experimental schemes to test the wave-particle duality in history: Young's two-slit interferometer and Mach-Zehnder two-arm interferometer. The perspective relies on design of two simultaneous sets of good measurement well separated in space, where one works to acquire the which-way information (particle behavior) and the other works to acquire the high-contrast interference pattern (wave behavior) separately, and they only make negligible perturbation to each other. We will discuss in detail how to make good measurement on the which-way information and interference pattern of quantum objects and analyze the effect of proper or improper arrangement and cooperation of these two sets of measurement on the outcome of wave-particle duality test.

**2. Analysis on Classical Interferometers for Wave-Particle Duality Test**

A standard Young's two-slit interferometer experiment is illustrated schematically in Fig. 1. This scheme is widely adopted to illustrate the wave nature of classical waves, such as water wave, sonic wave, and light wave, and it is used equally a lot to test the wave nature of quantum object. In both classical and quantum world, the interference of waves relies closely on the degree of coherence, which describes mathematically the mutual correlation status of wave in two space-time points. Physically, the interference pattern is directly related with the superposition of two secondary waves exiting the two slits of interferometer and transporting to the observation screen. The state of motion or transport of these waves can be uniformly described by the spatial function $\psi(\mathbf{r}, z)$. For optical waves, $\psi(\mathbf{r}, z)$ refers to the electric field (more precisely, one of its three vector components), which satisfies Maxwell's equations, while for massive particles such as electrons and atoms, $\psi(\mathbf{r}, z)$ refers to the probability wave function, which satisfies Schrödinger's equation. Yet, one important common thing is that in free space $\psi(\mathbf{r}, z)$ has a simple wave-motion (e.g., plane wave) form. Thus, in the following and all throughout this paper we use the term quantum object to represent all massless and massive particles, such as photons, electrons, and atoms.

When a wave with spatial distribution $\psi(\mathbf{r}, z)$ impinges upon the two slits of Young's interferometer, diffraction and interference takes places, and the pattern depends on the correlation between the wave state at the two slits ($\psi_{1,0}$ and $\psi_{2,0}$), which is just the way how the spatial coherence status of $\psi(\mathbf{r}, z)$ is defined.

Mathematically, in the paraxial approximation, the overall wave function $\psi_t(\mathbf{r},z)$ at the observation screen after the two-slit diffraction can be written as

$$\psi_t(\mathbf{r},z) = \psi_1(\mathbf{r},z) + \psi_2(\mathbf{r},z) = A(z)[\psi_{1,0}e^{i\varphi_1(\mathbf{r},z)} + \psi_{2,0}e^{i\varphi_2(\mathbf{r},z)}]. \quad (1)$$

Here $\psi_1(\mathbf{r},z)$ and $\psi_2(\mathbf{r},z)$ are the wave function evolved from the secondary source ($\psi_{1,0}$ and $\psi_{2,0}$) at the exit of the two slits, $\varphi_1(\mathbf{r},z)$ and $\varphi_2(\mathbf{r},z)$ are the phase accumulated during the wave transport process, and $A(z)$ is the single-slit diffraction factor. The signal intensity at the observation screen is then calculated by

$$I_t(\mathbf{r},z) = |\psi_t(\mathbf{r},z)|^2 = |A(z)|^2 [|\psi_{1,0}|^2 + |\psi_{2,0}|^2 + 2|\psi_{1,0}||\psi_{2,0}|\cos(\Delta\varphi)]. \quad (2)$$

Here the phase difference $\Delta\varphi(\mathbf{r},z) = \varphi_1(\mathbf{r},z) - \varphi_2(\mathbf{r},z)$ is the key factor for interference experiment, which determines the peak and valley position of the interference fringe. However, whether stable (temporal) and clear (spatial) interference pattern can be observed strongly depends on the temporal and spatial coherence nature of the wave function $\psi(\mathbf{r},z)$ and its values $\psi_{1,0}$ and $\psi_{2,0}$ at the two slits. This point has been well-established in classical optics, where to manifest good two-slit interference experiment, usually laser source is used to guarantee both high temporal and spatial coherence. When the coherence condition is well satisfied, perfect interference fringe pattern would be observed at the observation screen, following the formalism of Eq. (2).

Another experimental scheme popularly used for illustrating the wave nature of classical and quantum object is the Mach-Zehnder two-arm interferometer, as depicted in Fig. 2. A coherent optical beam or particle beam described by the wave function $\psi(\mathbf{r},z) = \psi_0(\mathbf{r})e^{i\varphi(z)}$ passes through the first beam splitter (BS 1) and is separated into two beams, which then transport along two paths (path $x$ and path $y$). Here $\mathbf{r}$ is the transverse position and $z$ is the transport position of the beam. In practice, the diffraction of beam is neglected so that $\psi_0(\mathbf{r})$ is a constant over $\mathbf{r}$ and the transport phase $\varphi(z) = kz$ (with $k$ being the wave number, and $z$ is the transport path) is the only relevant quantity. When the two beams pass through the second beam splitter (BS 2), interference occurs in both paths, and both detector *x* and detector *y* can record the periodic variation of the signal intensity when the path length difference of the two arms is changed continuously, remarking the wave nature of quantum object. Mathematically, the signal intensity in both arms is given by

$$I_t(\Delta\varphi) = |\psi_0|^2 [1+\cos(\Delta\varphi)]/2, \qquad (3)$$

where $\Delta\varphi$ is the phase difference of the two beams transporting across the two arms before finally reaching the detector *x* or *y*, and perfect 50:50 beam splitting has been assumed in both BS 1 and BS 2. Similar to the situation of Young's two-slit interferometer experiment, good temporal coherence of the beam $\psi(\mathbf{r},z)$ is the condition for this type of interference experiment.

The above two experimental setups are popularly used for exclusively illustrating the wave nature of photons and massive quantum objects. Simple modification can be made to both setups for exclusively revealing the particle nature of photons and massive quantum objects. In the Young's two-slit interferometer setup, one slit is closed to allow the beam only passing through the other slit. The transport path of particle is then determined while letting the interference fringe at the observation disappear, because now the signal intensity is $I_t(\mathbf{r},z) = |A(z)|^2 [|\psi_{1,0}|^2 + |\psi_{2,0}|^2]/2$. In the Mach-Zehnder interferometer, the BS 2 is removed, so that the path of the particle can be unambiguously determined by looking at the signal recorded by detector *x* and *y*. However, the signal observed by both detectors is now $I_t(\Delta\varphi) = |\psi_0|^2/2$, losing the periodic variation of signal and thus unable to tell the wave nature of the particle. Obviously the old and new version of the Young's and Mach-Zehnder interferometer each cannot simultaneously tell the wave and particle nature of the quantum object, instead they only tell either one or the other nature (i.e., wave or particle).

The above description is more or less based on conceptual argument. Strictly speaking, the behavior of optical and quantum particle beams in all these setups can be perfectly predicted by quantum theory itself. Some very simple elementary analysis and calculation already ~~only~~ can do the job well. The results in the old and new setup of interferometer are very different just because the environment within which the quantum object transports is very different. In some sense, quantum theory operates very well in these problems in the framework of its own methodology, but this methodology in itself just is not able to get people closer to the conceptual insight of the nature of quantum objects. However, just based on the above analysis (simple but classical, adopted by all people now and in history) for two classical interferometer setups, it is unfair to jump too hurriedly to the conclusion that quantum theory would completely exclude the possibility that the wave and particle nature of quantum object can be simultaneously observed by some very different while brand new experimental setups.

Notice that Eq. (2) is the basic formalism for describing the interference pattern in the observation screen that has been used almost exclusively in all Young's interferometers. Yet, in the context of quantum mechanics, there should exist a certain operation of measurement in reality (whether perfect or imperfect, at least in principle) that can well correspond to this formalism and whose outcome can be described by this formalism, otherwise the measurement is not working to allow for unambiguous acquisition of the interference pattern. In classical optics this formalism has been proven to be true based on a very solid physical mechanism of how optical interference pattern is observed: $I_t(\mathbf{r}, z)$ in Eq. (2) is just the optical intensity distribution pattern. Yet, great caution and care should be made for more general quantum particles. It might not be automatically true for arbitrary operation of measurement without deeply thinking of the corresponding physics underlying such a measurement. This point will become clear in later discussions.

It is worthwhile to examine and address various interferometer experiments for wave-particle duality and complementarity principle test through mathematical solution based on the orthodox operation methodology of quantum theory, instead of purely conceptual argument and justification. Solution of quantum mechanical problem for quantum objects transporting in these interferometers based on the Schrödinger equation (or Maxwell's equations) should rely on both the environment, which determines the Hamiltonian, and the initial condition of wave function (the coherence status). After the final wave function of quantum objects is solved, the outcome of the quantum system under various operations of observation and measurement can also be calculated and determined via standard formulation of quantum mechanics, such as perturbation theory and scattering theory used to handle the transition between initial and final quantum state. In this regard, the experimental setup for acquiring the wave behavior and particle behavior in the above standard Young's and Mach-Zehnder interferometers offers completely different environment for quantum objects to transport and evolve. The totally different and mutually exclusive outcome, i.e., either wave or particle, but never both, in these experiments are thus a natural result of quantum mechanics. If more complicated situations such as various "delayed-choice" experiments as first conceptually proposed by Wheeler are concerned [36-46], then the great details of how the environment (thus the Hamiltonian) changes in time and space must be first clarified and determined to allow for precise prediction of the outcome of these experiments. Usually a good physical model faithfully representing the experiment operation details needs to be placed into the Schrödinger equation in order to obtain good theory-experiment agreement. Unfortunately, this reasonable roadmap of problem solution (other than purely conceptual argument and deduction) was rarely adopted to reveal the puzzling wave-particle duality in these "delayed-choice" experiments.

## 3. New Design of Mach-Zehnder Interferometer

In the modified Mach-Zehnder's two-arm interferometer, the BS 1 remains in the path while BS 2 has been removed. It is well-known that this setup can unanimously reveal the particle nature of quantum object, but not the wave nature. Let us temporarily follow the transport path of quantum objects within the new Mach-Zehnder's interferometer rather than hurrying at once to the final destination of the path, the detector *x* and detector *y*. Notice that the initial coherent beam of quantum objects is described by the wave function $\psi(\mathbf{r},z) = \psi_0(\mathbf{r})e^{i\varphi(z)}$. However, in the above analysis made by ourselves (to be fair, also by all others now and in history), the transverse function $\psi_0(\mathbf{r})$ is completely omitted assuming its irrelevance. Yes, it is true that $\psi_0(\mathbf{r})$ is irrelevant to the final outcome of the detector *x* and detector *y*, either in the original or in the modified setup of Mach-Zehnder interferometer. Yet, don't forget that $\psi_0(\mathbf{r})$ is an indispensible part of the wave function $\psi(\mathbf{r},z)$ and it also carries the information of coherence of the beam. This information would manifest itself in the cross region of the two beams in space. It is worth making a more mathematical analysis on the behavior of wave transport in this region.

For simplicity, both beams are approximately modeled as plane wave represented by wave vector $\mathbf{k}_x = (k_0, 0, 0)$ for the horizontal beam (path *x*) and $\mathbf{k}_y = (0, -k_0, 0)$ for the vertical beam (path *y*). Then the wave function in the cross region is given by

$$\psi(x,y) = \psi_0(e^{ik_0 x} + e^{-ik_0 y + i\Delta\varphi})/\sqrt{2}. \qquad (4)$$

The corresponding wave intensity is simply

$$I(x,y) = |\psi(x,y)|^2 = |\psi_0|^2 [1 + \cos(k_0 x - k_0 y + \Delta\varphi)]. \qquad (5)$$

Obviously a volume interference pattern forms in the crossing region of the two coherent beams. Therefore, the general picture of this new Mach-Zehnder interferometer is that the two quantum-object beams transport along their own path, cross and interfere, leave each other without changing their original paths, and then finally reach their own destination, detector *x* and detector *y*, respectively.

When summing up all the above information and thinking over from a new angle of conceptual view rather than the standard one, it seems natural for us to immediately point to a very strong conclusion that the modified setup of Mach-Zehnder interferometer, as illustrated in Fig. 3, can simultaneously record the wave and particle nature of quantum objects, although in different temporal and spatial domains by using two different sets of measurement instrument at the same time, one in the two-beam cross region for acquiring the wave nature, the other in the destination for acquiring the particle nature. Having said this, we still need to go one step further to make the principle idea of the above gedanken experiment operational practically and leave no loop holes for doubt. The key then is to design a smarter instrument to observe the interference pattern in the two-beam cross region while without significantly modifying the environment of quantum object transport paths. In optics, this is not difficult to do. One can place a glass plate coated with a very thin (say, a few nanometers thick) scattering or absorption material with its surface inclined in some angle (say $30^\circ$) with the *x*-axis. In the scheme of weak scattering, the scattering intensity by the thin film would be proportional to the intensity profile of light at the surface, which can then be collected by an optical telescope outside the path of Mach-Zehnder interferometer. In the absorption scheme, the interference pattern is directly recorded by the thin film. In both schemes, the scattering or absorption must be sufficiently strong to allow the interference pattern to be recorded by state-of-the-art sensing materials or devices with very high peak-valley contrast, and at the same time must be sufficiently weak so that most photons (say >99%) can transmit through the thin film and follow their original path to reach the detector for its particle nature faithfully recorded. In both weak scattering and absorption schemes, the measured optical pattern is proportional to the spatial interference distribution and thus has a perfect peak-to-valley contrast in ideal conditions.

Notice that the above description on the operation mechanism of the weak detection scheme for the interference pattern not only works well in the framework of classical optics for light waves, but also works equally well in the framework of quantum mechanics for photons. The reason is the same as for the beam splitter used in this interferometer and commonly used in many other quantum systems. It is well known that the functionality of a beam splitter, such as the amplitude and phase of the two split beams, can be designed and described exactly based on the knowledge of classical physical optics. Yet, it equally works for a single photon, where the probability amplitude and phase of the wave function for the two split beams of photon are the same as those designed or predicted by classical optics. In fact, many other devices involved in quantum systems also work exactly in the way described by classical physics, because they are essentially classical. This means that the above weak measurement scheme for interference pattern observation, although designed and discussed basically in the context of classical physics, can be described equally well quantum mechanically and should yield equal experimental results followed not only by classical light waves, but also by an ensemble of photons, or equivalently by a single photon in the context of probability.

The modified Mach-Zehnder interferometer designed to manifest both the wave and particle nature simultaneously for quantum objects can work not only in principle (in gedanken experiment) but also in practice (using modern technology), and it can be immediately brought to experimental test for photons. As to other massive particles such as electrons and atoms, the key to successful experimental demonstration of their new wave-particle duality also relies in design and fabrication of instruments or materials like the above thin film for photons that allow for weak measurement of the interference pattern in the two-beam cross region. The current scheme is drastically different from many previous schemes based on Mach-Zehnder interferometer used for wave-particle duality test. Previous schemes, like Wheeler's delayed-choice gedanken experiment scheme, try to operate on each single quantum object to reveal both the wave and particle nature, however, two sets of instrument are used (namely, inserting BS 2 or removing BS 2) but not simultaneously, which obviously repel each other. In the methodology of quantum theory, these two sets of instrument arrangement correspond to two mutually repulsive quantum mechanical problems (corresponding to two different setups of Mach-Zehnder interferometer), whose solution can never lead to simultaneous wave and particle observation no matter how clever the two sets of measurement are arranged in space-time. In comparison, the current scheme only uses a single set of measurement involving two instruments simultaneously placed in different spatial and temporal domains without mutual strong influence, thus it allows for simultaneously observing the wave and particle natures of quantum objects without offending any standard methodology of quantum theory.

## 4. Young's Interferometer for Quantum Objects with Internal Freedoms

In the history of quantum theory, Young's two-slit interferometer has become a popular gedanken experiment used to debate and clarify the concept and philosophy of quantum mechanics, among which Einstein's recoiling slit [3], Heisenberg's $\gamma$-ray microscope [4], and Feynman's light-electron scattering scheme [12] are several prominent examples. The failure to obtain simultaneous definite information of wave and particle by using these old experiment setups is closely related with the Heisenberg's uncertainty principle. The act of measurement on which-way information inevitably disturbs the particle in such a manner that coherence is lost and the interference pattern disappears. Mathematically, the wave function diffracted by the two slits at the presence of strong perturbation can be written as

$$\psi_t(\mathbf{r},z) = \psi_1(\mathbf{r},z)e^{i\delta\varphi_1} + \psi_2(\mathbf{r},z)e^{i\delta\varphi_2} = A(z)[e^{i\delta\varphi_1}\psi_{1,0}e^{i\varphi_1(\mathbf{r},z)} + e^{i\delta\varphi_2}\psi_{2,0}e^{i\varphi_2(\mathbf{r},z)}], \quad (6)$$

under the simplified model where the perturbation induces a large random phase $\delta\varphi_1$ and $\delta\varphi_2$ to each secondary wave function from the two slits. Then the signal recorded at the observation screen is given by

$$\begin{aligned} I_t(\mathbf{r},z) &= |A(z)|^2 \, [|\psi_{1,0}|^2 + |\psi_{2,0}|^2 + 2|\psi_{1,0}||\psi_{2,0}|\langle e^{-i\delta\varphi_1}\rangle\langle e^{i\delta\varphi_2}\rangle\cos(\Delta\varphi)] \\ &= |A(z)|^2 \, (|\psi_{1,0}|^2 + |\psi_{2,0}|^2). \end{aligned} \quad (7)$$

Here the bracket $\langle...\rangle$ means ensemble average. Obviously, the methodology of quantum theory once again successfully predicts the outcomes of the strongly perturbed Young's two-slit interferometer gedanken experiments. On the other hand, if the which-way measurement does not induce a large random phase shift, but induces a constant phase shift (or a random but negligibly small phase shift) for every particle, then Eq. (2) still holds and the which-way information acquisition operation does not destroy the interference pattern. Of course, to achieve this goal in practice rather than just in thought experiment, the observation induced perturbation must be sufficiently weak so as to maintain the spatial coherence of the two-slit secondary waves, as understood in the language of wave optics or wave mechanics.

To this end, Scully and coworkers have proposed a modified scheme of Young's two-slit interferometer involving a micromaser cavity which-way detector to interact with the internal state of long-lived Rydberg atoms [28,29]. Although the which-way detector only induces negligibly small while random phase shift to the center-of-mass wave function of atom [30-33], the authors conclude that their scheme still cannot acquire both the wave and particle information simultaneously due to the entanglement between the which-way detector and the atomic internal states [28,29]. Later experiments seem to justify this scheme of entanglement upholding complementarity principle [30,31]. In the following, we will revisit and generalize this type of gedanken experiment and make an elementary quantum mechanical analysis on the outcome under various situations from a new angle of conceptual view, which eventually leads to a totally different conclusion from the popular one in history.

The first thing that needs to do before going into the details of discussing wave-particle duality in an atomic system is to clarify the object of examination. This could become a problem because atom itself is not an elementary particle, but rather a very complicated composite system. So it is the motion of the atom itself, more precisely, the atom center-of-mass, that should be considered as the most appropriate representation of the atom itself as a quantum object. As is well-known, an atom always possesses some internal structures and freedoms. For example, it consists of a nucleus at its center and some electrons around the nucleus. The nucleon consists of many protons and neutrons, which in turn are made up of quarks. When observed, the atom might interact with the environment (e.g., which-way detectors) through some of its internal freedoms during the process of its center-of-mass motion. It is then obvious that when we talk about the interference of atoms, we should only refer to the atom center-of-mass itself, rather than the interference of any part of these internal composites. Otherwise, ambiguity inevitably occurs, and more seriously, incoherence in any internal freedom (such as electrons and quarks) will force one to conclude that the atomic wave cannot be coherent, which is of course not the case. Essentially, the effect of internal freedoms in an atom interferometer should be reflected only through their influence on the atom center-of-mass motion, otherwise, they are just opaque to the operation of atom interferometer.

The motion of atom center-of-mass can also be described by quantum state $\psi(\mathbf{r}, z)$. In free space, $\psi(\mathbf{r}, z)$ can be approximately described by plane wave $\psi(\mathbf{r}, z) = \psi_0 e^{ikx}$ when high-coherence atomic beam is concerned, and the atom center-of-mass momentum is given by $p = \hbar k$. When only considering Young's two-slit interferometer for the atom center-of-mass, the formalism [Eqs. (1)-(3)] applies equally well, and the coherence of atomic beam is the key to observing high-quality interference pattern. However, it is also unable to acquire both wave and particle information simultaneously using the conventional scheme where the which-way detector directly works on the atom center-of-mass. Things might become very different when the internal freedom of atom is adopted to interact with the which-way detector and thus serves as the probe of which-way information.

In the new Young's two-slit interferometer setup for quantum objects with internal freedoms described by quantum state $\varphi$, the total wave function of the atomic system after passing through the two-slit and being subject to diffraction and interference effect is given by

$$\Psi(\mathbf{R}) = \psi_1(\mathbf{R})\varphi_1 + \psi_2(\mathbf{R})\varphi_2. \qquad (8)$$

If one directly uses the methodology of quantum theory literally without going into details of its physical meaning, the interference pattern is given by

$$I(\mathbf{R}) = |\Psi(\mathbf{R})|^2 = |\psi_1(\mathbf{R})|^2 + |\psi_2(\mathbf{R})|^2 + 2\operatorname{Re}[(\varphi_1^*\varphi_2)\psi_1^*(\mathbf{R})\psi_2(\mathbf{R})]. \qquad (9)$$

This formulation has been overwhelmingly and popularly adopted in analysis of atom interferometers, and it implies that the interference fringe pattern is modulated by both the center-of-mass wave motion (denoted by coordinate $\mathbf{R}$) and the internal quantum states. Here we have assumed that $\varphi_1$ and $\varphi_2$ are normalized wave function defined in a certain Hilbert sub-space so that $\langle \varphi_1 | \varphi_1 \rangle = 1$ and $\langle \varphi_2 | \varphi_2 \rangle = 1$. Yet, Eq. (9) needs more careful examination. As usually in practice, $\varphi_1$ and $\varphi_2$ are quantum states in their specific Hilbert sub-space, rather than in the usual real space, unlike the center-of-mass motion state $\psi_1(\mathbf{R})$ and $\psi_2(\mathbf{R})$. Therefore, it seems hard to interpret the total state in Eq. (8) according to the probability definition of orthodox quantum mechanics in terms of Born's interpretation of wave function for quantum state, namely, the probability of finding a particle at position $\mathbf{R}$. Similarly, the definition of the interference pattern of the atom center-of-mass according to Eq. (9) is also problematic. This issue will make the wave-particle duality more complicated to describe and harder to imagine and understand. That this issue can have very significant consequence will become clear from the following examples.

In general experiment of wave diffraction and interference of quantum objects, the internal quantum state does not play an active role, in other words, the internal quantum state is not intentionally prepared in some specific states. This means that the quantum states in the two slits are generally not coherent so that $\langle \varphi_1^* \varphi_2 \rangle = \langle \varphi_1^* \rangle \times \langle \varphi_2 \rangle = 0$. Then according to Eq. (9), $I(\mathbf{R}) = |\psi_1(\mathbf{R})|^2 + |\psi_2(\mathbf{R})|^2$ and the interference fringe completely disappears as the internal state modulation amplitude is zero. If special care is made to prepare the internal quantum state to be highly coherent, then Eq. (9) holds true, but whether the interference pattern appears or not sensitively depends on the value of $\langle \varphi_1 | \varphi_2 \rangle$. If the two internal states are orthogonal, then the interference pattern disappears but leaving the possibility to acquire the which-way information by looking at the internal state. Otherwise, if $\varphi_1 = \varphi_2$, then the interference pattern appears but making it impossible to acquire the which-way information.

Yet, when the above conceptual argument and methodology operation extend without any constraint to all the internal states of electrons and nucleus comprising the atom, then all these internal states cannot be coherent simultaneously. If we apply Eq. (9) literally, then no interference pattern can be observed in any Young's two-slit interferometer for composite quantum objects with internal freedoms, even if one does not trouble to observe simultaneously the which-way information for the quantum object transporting in the interferometer. This of course is not the case as numerous experiments in history have successfully demonstrated the wave nature of quantum objects through diffraction and interference. So something must be wrong physically in directly and literally applying the orthodox definition that the interference pattern is simply given by $I(\mathbf{R}) = |\Psi(\mathbf{R})|^2$, even if $\Psi(\mathbf{R})$ is the total wave function for the composite quantum system involving multiple quantum states defined in various Hilbert sub-spaces.

## 5. The Role of Which-Way Detectors Interacting with Quantum Objects

Things will become even more complicated when the which-way detectors are further introduced to form an even larger coupled quantum system together with the atomic system. Through direct and literal application of $I(\mathbf{R}) = |\Psi(\mathbf{R})|^2$ to define the interference pattern, many conceptual conclusions can be and indeed have been deduced. Although obviously counter-intuition and seemingly unphysical in many situations, defenders and proponents of the orthodox Copenhagen interpretation of quantum theory would introduce more and more new and fashionable concepts and methodologies such as entanglement, delayed-choice, quantum eraser, just to name a few important, to argue that these conceptual conclusions reflect the truth in quantum world. A prominent example would be the micromaser which-way detector gedanken experiment proposed and analyzed by Scully [28,29]. A simplified version of the gedanken experiment is illustrated in Fig. 4, where a coherent beam of Rydberg atom with excited internal state passes through two micromaser cavities each placed just ahead of one of Young's two slits. The micromaser cavity serves as the which-way detector via interacting with the internal quantum states of atom.

Let us tentatively follow Scully by going beyond the atom itself and considering simultaneously the detector. In another word, we now need to look at the evolution of the atom-detector as a whole system. Suppose the quantum state of the two which-way detectors is described by $\phi_1$ and $\phi_2$ in another Hilbert sub-space, respectively, corresponding to atom passing through either slit. In addition, assume that the detector only interacts with the internal quantum state of the particle and neglect the direct interaction of the detector with the center-of-mass motion of the

atom. Then the total wave function of the coupled atom-detector quantum system is given by

$$\Psi(\mathbf{R}) = \psi_1(\mathbf{R})\varphi_1\phi_1 + \psi_2(\mathbf{R})\varphi_2\phi_2. \tag{10}$$

Again one uses the operation methodology of orthodox quantum theory and finds that the interference pattern is given by

$$I(\mathbf{R}) = |\Psi(\mathbf{R})|^2 = |\psi_1(\mathbf{R})|^2 + |\psi_2(\mathbf{R})|^2 + 2\operatorname{Re}[(\varphi_1^*\varphi_2)(\phi_1^*\phi_2)\psi_1^*(\mathbf{R})\psi_2(\mathbf{R})]. \tag{11}$$

When the internal quantum stat of atom $\varphi$ interacts with the micromaser cavity quantum state $\phi$ so that the which-way information of atom is unambiguously transferred to and recorded by the which-way detector, one has relation that $\varphi_1 = \varphi_2 = \varphi$ and $\langle \phi_1 | \phi_2 \rangle = 0$, thus the interference term in Eq. (11) $\operatorname{Re}[(\varphi_1^*\varphi_2)(\phi_1^*\phi_2)\psi_1^*(\mathbf{R})\psi_2(\mathbf{R})]$ is equal to zero. This means that the unambiguous recording of the which-way information in the Young's interferometer will definitely sweep away the interference pattern for the atomic center-of-mass motion. This thought outcome would thus seemingly uphold the principle of complementarity.

In the current modified two-slit interferometer the acquisition of the which-way information does not suffer from Heisenberg's uncertainty principle induced by uncontrollable interaction of the which-way detector with the atomic internal freedoms, because the interaction energy is much smaller than the center-of-mass momentum energy [28-33]. This which-way detector seems to be smarter than the old ones such as Einstein's recoiling slit [3]. In the scheme of Einstein's recoiling slit, one needs to directly operate on the particle center-of-mass motion in order to acquire the which-way information, and the interaction length scale must be comparable with the slit width, leading to an interaction sufficiently strong to induce uncontrollable perturbation upon the center-of-mass motion and destroy the interference pattern. Obviously it is the Heisenberg's uncertainty principle that works to greatly deteriorate the coherence of the atomic beam. In contrast, in the scheme of Scully's micromaser interferometer, the interaction length scale of the micromaser with the atomic internal freedoms is on the scale of the wavelength of microwave photon and comparable with the cavity size, which could be much larger than the slit width. Then the emission of microwave photon and the subsequent recoil of atom only induce negligible influence to the coherence status of atomic center-of-mass. However, it was argued and generally agreed that the entanglement between the which-way information and the micromaser quantum state would enforce a destructive factor in the atom-detector quantum state as described in Eq. (11) to sweep away the interference pattern [28,29]. Notice that these arguments about the entanglement enforced complementarity principle are more or less based on the confidence and belief that the conceptual basis for orthodox Copenhagen interpretation of quantum theory, the principle of complementarity is always correct no matter how long time goes and how much technologies develop.

In some sense, the analysis is like looking for supporting materials based on an *apriori* conceived conclusion, i.e., the complementarity principle is always true, rather than deducing conclusion from supporting materials in normal logics. The logic way of argument about the role of entanglement in the wave-particle duality is counter-intuition for many peoples. Yet, many further arguments and outcomes as a consequence of this analysis, including the concept of delayed-choice experiment and quantum erasers looks even more counter-intuitive and harder to swallow for peoples who are not the hot-faith proponents of the orthodox conceptual quantum theory [5-8,11-18,47]. Being a realist rather than a doctrinaire of quantum theory, I try to make a more comprehensive analysis of such a type of gedanken experiment following the well-established regular methodology of quantum theory and using regular logics, in a hope to offer an alternative and common-sense solution to this old and classical puzzling problem of wave-particle duality. It is found that the key lies in how to measure the interference pattern.

**6. Analysis on Interference Pattern Measurement**

The above discussions largely focus on how to acquire the which-way information of atomic system. The Scully scheme has offered at least in principle a smarter way to achieve this goal without suffering from the constraint of Heisenberg's uncertainty principle. Yet, there is another issue that is equally important in the wave-particle duality test by using atom interferometer, namely, the acquisition of interference pattern. Obviously relatively much less attention has been paid to handling this issue not only in the Scully work but also in many other gedanken or practical experiments in history. In some sense, the situation of much less intensive and extensive analysis, or even negligence and ignorance, on the great details of measurement process of the interference pattern formed by quantum objects in wave-particle duality test might be the reason why so many concepts and interpretations that are apparently counter-intuition, seemingly unphysical or metaphysical, but are still overwhelmingly thought to be correct quantum mechanically, are raised and more or less cause great confusion in conceptual understanding of these gedanken or practical wave-particle duality test experiments. Thus, it is invaluable to make a detailed analysis on how to measure the interference pattern in various interferometers.

First, let us go to the simplest situation of quantum object without the internal freedom. The virtual interference pattern of two-slit interferometer is given by Eqs. (1) and (2). I use the term of "virtual" here because the interference pattern at this step is largely only of mathematical or conceptual meaning. To bring it into reality, some certain experimental technique or measurement (a term popularly used by the proponents of orthodox quantum theory) must be performed to transfer the "virtual" probability distribution profile of quantum object into a form that can be sensed by human beings or recorded through some instruments familiar with human beings. However, the paramount principle is that the measurement should faithfully reflect the "virtual" profile. In optics, a simple flat rough-cast screen or glass can be used to record this interference pattern as the scattering intensity of light off the screen as visualized by human eyes is just proportional to the intensity profile at the screen, thus the "virtual" interference pattern is recorded by the screen. In atomic optics, position sensitive detector can be used to record the distribution of atom. On the other hand, the measurement device should not deviate much from this linear operation regime, otherwise, the "virtual" interference pattern would not be faithfully recorded. In optics, a black surface obviously cannot implement such a recording functionality as it smears off all patterns irrespective of their profiles.

Why it is important to talk about the measurement process for interference pattern will become clear when it comes to the Young's interferometer for composite quantum objects with internal freedoms. A good example is the Rydberg atom beam with its internal quantum states interacting with the micromaser cavity which-way detector, as discussed in Fig. 4. The total wave function for the composite quantum object when reaching the observation screen is given by Eq. (8) as $\Psi(\mathbf{R}) = \psi_1(\mathbf{R})\varphi_1 + \psi_2(\mathbf{R})\varphi_2$. Only $\psi_1(\mathbf{R})$ and $\psi_2(\mathbf{R})$ are related with the center-of-mass motion of the quantum object in real space. They are subject to phase shift depending on their transport path and responsible for the formation of interference pattern. Of course, this is just "virtual" interference pattern. On the other hand, the quantum state for internal freedoms $\varphi_1$ and $\varphi_2$ are defined in another Hilbert sub-space. They will evolve by themselves, say, having phase shift or others, during their transport accompanying with their master quantum object. Usually the center-of-mass motion and the internal freedoms are independent of each other during the transport of quantum object in free space, unless indirect interaction happens when either one freedom or both are subject to some external perturbation.

The paramount principle of measurement should also be followed to faithfully record the interference pattern, now for the center-of-mass of quantum object. Frankly speaking, this is not always an easy task that any measurement instrument would automatically work to implement. In contrast, it requests great caution and care so as not to induce ambiguity, confusion or even error while at the same time being not aware of them consciously. I would not reexamine how measurement of the interference pattern was made in all the wave-particle duality gedanken and practical experiments performed in history, as this seems to be an impossible task. Instead, I would argue and discuss the principle of how to make such a good measurement considering various situations.

Generally the measurement can be modeled as an operation $M$ working on the wave function of the composite system $\Psi(\mathbf{R}) = \psi_1(\mathbf{R})\varphi_1 + \psi_2(\mathbf{R})\varphi_2$. The new wave

function after the measurement process is then

$$\Psi_M(\mathbf{R}) = M\{\Psi(\mathbf{R})\} = M\{\psi_1(\mathbf{R})\varphi_1\} + M\{\psi_2(\mathbf{R})\varphi_2\}. \tag{12}$$

In general situations, the operator $M$ could work either on the center-of-mass motion, or on the internal freedom, or on both. In the first case, $M$ only work on $\psi_1(\mathbf{R})$ and $\psi_2(\mathbf{R})$, then the internal quantum state does not participate in the measurement process, or in other words, they are opaque to the measurement. In this case the composite quantum object behaves the same as a pure quantum object. Then Eq. (12) transforms to

$$\Psi_M(\mathbf{R}) = M\{\Psi(\mathbf{R})\} = A[\psi_1(\mathbf{R}) + \psi_2(\mathbf{R})]. \tag{13}$$

Here a good measurement has been assumed where the measurement only introduces a linear amplitude factor. Besides, the measurement should also yield a signal intensity given by

$$I_M(\mathbf{R}) = |\Psi_M(\mathbf{R})|^2 = |A|^2 \{|\psi_1(\mathbf{R})|^2 + |\psi_2(\mathbf{R})|^2 + 2\operatorname{Re}[\psi_1^*(\mathbf{R})\psi_2(\mathbf{R})]\}. \tag{14}$$

Obviously a perfect interference pattern representing the positioning probability of the quantum object center-of-mass has been recorded by this good measurement, which is the term $\operatorname{Re}[\psi_1^*(\mathbf{R})\psi_2(\mathbf{R})]$. Of course, in practice there are many other possible bad measurements that will deteriorate the interference pattern by either introducing random phase shift, or imposing nonlinear or more complicated modulation to $\psi_1(\mathbf{R})$ and $\psi_2(\mathbf{R})$, although these measurements also neglect completely the internal freedoms. These bad measurements should be avoided in practical experiments.

In the second case, the measurement operator $M$ only works on the internal freedom and induces transition of $\varphi_1$ and $\varphi_2$ to new states $\varphi_{1,M}$ and $\varphi_{2,M}$. This scheme has been adopted in most atom interferometers for wave-particle duality test, such as those reported in Ref. [34,35]. For instance, light emission when atom jumps from a higher electronic state to a lower electronic state under the dipole transition can be used to probe the position of atom by recording the emission site optically. In this situation, Eq. (12) transforms to

$$\Psi_M(\mathbf{R}) = M\{\Psi(\mathbf{R})\} = \psi_1(\mathbf{R})\varphi_{1,M} + \psi_2(\mathbf{R})\varphi_{2,M}. \tag{15}$$

It is seen that the center-of-mass wave function just imposes a position dependent probability modulation to the internal state transition. Suppose the recorded signal intensity is proportional to the transition probability of quantum state, as is the case in light emission experiment. The signal is given mathematically as

$$I_M(\mathbf{R}) \propto \langle \Psi(\mathbf{R}) | M | \Psi_M(\mathbf{R}) \rangle = |\psi_1(\mathbf{R})|^2 \langle \varphi_1 | M | \varphi_{1,M} \rangle + |\psi_2(\mathbf{R})|^2 \langle \varphi_2 | M | \varphi_{2,M} \rangle \\ + \psi_1^*(\mathbf{R})\psi_2(\mathbf{R})\langle \varphi_1 | M | \varphi_{2,M} \rangle + \psi_1(\mathbf{R})\psi_2^*(\mathbf{R})\langle \varphi_2 | M | \varphi_{1,M} \rangle. \tag{16}$$

Obviously the best way to record a perfect interference pattern with

$I_M(\mathbf{R}) \propto |A|^2 \{|\psi_1(\mathbf{R})|^2 + |\psi_2(\mathbf{R})|^2 + 2\operatorname{Re}[\psi_1^*(\mathbf{R})\psi_2(\mathbf{R})]\}$ is to make

$\langle \varphi_1 | M | \varphi_{2,M} \rangle = \langle \varphi_2 | M | \varphi_{1,M} \rangle$, and the simplest way to satisfy this is that $\varphi_1 = \varphi_2$ and $\varphi_{1,M} = \varphi_{2,M}$. On the other hand, if $\langle \varphi_1 | M | \varphi_{2,M} \rangle = \langle \varphi_2 | M | \varphi_{1,M} \rangle = 0$, then the interference pattern disappears even if the measurement imposes no uncontrollable perturbation to the quantum object center-of-mass. The analysis clearly shows that the interference pattern obtained by this type of measurement is the convolution of the internal state transition intensity and the quantum object center-of-mass positioning probability. Both factors have equally important contributions to the apparent interference pattern. Strictly speaking, this composite-measurement scheme would be subject to serious ambiguity as it strongly depends on the internal state evolution of atom under probe, and essentially it is also the origin of many delayed-choice outcome of wave-particle duality test, like those presented in Ref. [34,35]. This point should be kept in mind when revisiting the wave-particle duality test experiments (either gedanken or practical) in history.

In the above, we discuss how to make a good measurement upon the interference pattern formed by quantum objects without internal freedoms and with internal freedoms. As the two quantities exactly overlap in space, probe over them can both reveal the interference feature of the quantum object, albeit caution and care should be taken. On the other hand, the probe over the which-way information is made by another set of measurement instrument that is separated in space from the interference measurement instrument. These two sets of measurement could be independent, as a result of which it is then possible to acquire both the which-way information and interference pattern simultaneously. However, in many examples of wave-particle duality test experiments, these two sets of measurement are assumed to be closely correlated or entangled even if they are separated in space, rendering it impossible to acquire both the wave and particle behavior simultaneously and upholding the consistency of complementarity principle [28,29].

Let us take a closer look at the Scully gedanken experiment [28,29] and focus on the measurement process. The total wave function of the coupled atom-detector quantum system is given by $\Psi(\mathbf{R}) = \psi_1(\mathbf{R})\varphi_1\phi_1 + \psi_2(\mathbf{R})\varphi_2\phi_2$ when the atom just passes through the two slits. This type of wave function represents a typical entanglement state and it is the atom-detector interaction that creates this entanglement. After the interaction ends, the two sub-systems evolve by themselves without disturbing each other and each can be subject to measurement. The atom, with its center-of-mass carrying internal quantum state, transports in free space and reaches the observation screen where the measurement over the interference pattern is implemented. The micromaser cavity will eventually have the emitted microwave photon by the transmitted atom absorbed by its wall and fulfill the measurement of the which-way information by counting the photon number (0 or 1) every time in the two cavities. If we generalize the atom-detector system as a quantum object (atom) with internal freedom (the state of which-way detector), then according to the above discussion on measurement process, one can simply adopt a measurement that either operates directly on the atom center-of-mass motion or indirectly operates on the atom internal state (which imposes no trouble as $\varphi_1 = \varphi_2 = \varphi$), from which it is not difficult to acquire the interference pattern, either appearing or disappearing. In either way, the signal intensity at the screen is given by Eq. (14), where nearly perfect interference pattern should be observed as $\psi_1(\mathbf{R})$ and $\psi_2(\mathbf{R})$ nearly resemble (more precisely, with addition of only a negligibly small random phase factor) the perfect function without interaction between the atom and the maicromaser cavity. On the other hand, because the well separation of the screen from the micromaser cavity, the measurement of the interference pattern will have no influence upon the which-way information measurement, i.e., the cavity photon number counting. In this way, the wave and particle behaviors of the atom could be simultaneously illustrated by using this experiment scheme.

If, on the other hand, a very extraordinary, and strange in some sense, measurement operation is made to examine the wave-particle behavior in the coupled atom-system, things can be very different and favor the upholding of complementarity principle. As depicted in Fig. 4, if one modulates the atom center-of-mass wave function (exhibiting interference pattern) with the outcome of a special measurement made upon the internal freedom of atom (the micromaser cavity state), which is just a simple coincidence count measurement on the photon number in the two cavities, then this modulation factor is exactly zero, destroying the interference pattern as implied by Eq. (11). Mathematically, this coincidence counting measurement is equal to determine the term $\langle \phi_1 | \phi_2 \rangle$, which should be exactly zero as expected. In some sense, this coincidence counting is a good measurement for which-way information acquisition, but it is a bad measurement for interference pattern observation when it is inappropriately used to modulate the interference pattern. The cooperation of the good and bad measurement leads to the failure of simultaneous illustration of wave and particle nature in this special Young's two-slit interferometer and upholding of the complementarity principle of the orthodox quantum theory. Although the two sets of measurement are mutually exclusive (either wave or particle, but not both), they are just two very special cooperated measurements among so many possible operations of measurement that can work on the current two-slit interferometer. One can easily go around them and take other good measurements for illustrating both natures simultaneously, like the ones that have been mentioned earlier. Yet, if one wishes to uphold the complementarity principle and the conceptual orthodox Copenhagen interpretation of quantum theory, then the two special sets of measurement in cooperation via artificial data processing is the only way to achieve this goal, no matter how counter-intuition, extraordinary, and even absurd such an action might looks. However, such an action necessarily leaves apparent loophole of conceptual justification.

The inappropriate and artificial connection of the which-way detector measurement and the interference pattern measurement as illustrated in Fig. 4 can be be used to explain the outcome of so many delayed-choice experiments and quantum eraser experiments that have been further deduced to be intrinsically involved and manifested in this type of interferometer [36-46]. The reason is that one can now connect whatever operation on the which-way detector (instant or delayed-choice) with the distant interference pattern measurement operation and generate any outcome of experiment, removing or restoring the interference pattern. In some sense, this connection is not a physical one that instantly occurs in the interferometer but just a mathematical operation that can be arbitrarily manipulated in a way according to one's subjective desire via post management and processing (e.g., convolution and modulation) of two sets of measurement data and can be implemented at arbitrary time long after the experiment, not to mention the usual delayed choice.

**7. Further Discussions**

The above discussions show that the key to acquire both the wave and particle information for quantum objects in a good interferometer lies in the simultaneous use of two sets of good measurement that are well separated in space and do not perturb each other. In addition, all the analyses are made based on elementary methodology of quantum mechanics rather than abstract and counter-intuition conceptual arguments and justifications. These gedanken experimental schemes can be readily tested by using modern quantum optics or atomic physics technologies [27,41-46,49-51].

The wave nature of quantum object manifests in the appearance or disappearance of high peak-valley contrast interference pattern, and the formation of this pattern requires an ergodic ensemble of quantum object to participate in either via time accumulation or via sample accumulation. The key in experiment is the maintenance of the coherence status of the ensemble. Only a single quantum object passing through an interferometer cannot illustrate a sufficient fine interference pattern in experiment.

This point should hold true in both the Mach-Zehnder interferometer as discussed in Sec. 3 and Young's interferometer as discussed in Sec. 4-6. Actually, such an assumption and concept have been justified in numerous experiments in history and current times that were performed to test the wave nature of massive particles such as electrons and massless particles such as photons. For instance, in experiments of particle diffraction and interference, the observed pattern always gradually evolves from some randomly kicking points in the observation screen for the initial one, two, three, and small numbers of particles and events, to final stable and well-defined diffraction or interference pattern for a sufficiently large number of particles and events. This feature can be well understood by the probability nature of wave function used to describe the motion of these particles in orthodox quantum mechanics. Therefore, any experimental setup, including the interference pattern detector screen in the Mach-Zehnder interferometer as illustrated in Fig. 3, that is used to observe the interference pattern of quantum object, must operate and should only operate in such a probabilistic way for an ensemble of particles in practical operation of measurement. In history and current times, nobody has ever designed or even imagined a technological means to demonstrate and observe the interference pattern for only one single quantum object.

On the other hand, the particle nature of quantum object manifests in the which-way information detection of each individual particle. This can be easily done by a wide variety of modern technologies. The difficulty here is to introduce as small as possible perturbation to the particle passing through the two slits and uphold the coherence status of the ensemble of particles as a whole. In history many schemes of interferometer failed to pass such a double-fold criterion. But now the internal freedoms of quantum object, which is intrinsic almost to all quantum particles, can be exploited to probe the which-way information of every particle while at the same time uphold the high coherence status of the whole ensemble of particles center-of-mass in various interferometers. In other words, it is now possible to design good schemes of measurement to acquire both the particle and wave nature of quantum object. Although this would seem to violate the popular wave-particle duality and complementarity principle that stand at the center conceptual stage of orthodox quantum theory, it strictly follow the orthodox methodology of quantum theory about the probability nature of wave function and its evolution following Schrödinger's equation either under a fixed environment or under interaction with detectors.

Another interesting issue is whether it is possible to experimentally test the wave-particle duality for an atomic beam with different internal freedoms transport within the Young's two-slit interferometer. In the Scully scheme, both the incident atomic beam going into the micromaser cavity (excited state) and the transmission beam passing through the two slits (lower state) have the same quantum states of internal freedom in the two paths. When the atomic beam reaches the screen, the internal freedom is allocated at the same quantum state of $\varphi_1 = \varphi_2 = \varphi$, therefore, one can directly operate on the atomic center-of-mass motion or the internal quantum state to measure the interference pattern. The which-way information is acquired via another measurement in the micromaser cavity. Yet, if the atomic beam passing through the two slit is in different internal quantum state, say $\varphi_1 \neq \varphi_2$, which can be done by only placing a micromaser cavity before the upper slit while leaving the lower slit open, then according to Eq. (12), one can adopt a measurement scheme that operates only on the atomic center-of-mass motion and misses the internal freedom to acquire the interference pattern of the ensemble of quantum object.

The schematic setup of this new Young's two-slit interferometer can be envisioned by only making a small change to the setup illustrated in Fig. 4. The which-way information can still be acquired by the single micromaser cavity by looking at whether there is emission of photon in the upper cavity. The fact that whether or not the atom passes through the lower slit cannot be known directly does not impose a trouble to change the outcome because logically if one knows the atom passes (or does not pass) the upper slit, then it must does not pass (or passes) the lower slit. There is something that one would know without making direct measurement, but just through many physical laws, such as energy conservation, momentum conservation, or particle number conservation, which suffice to serve the role of an indirect partial measurement. On the other hand, if one can find an even smarter scheme of measurement that allows for simultaneous determination of not only the atom center-of-mass position, but also its internal state (either $\varphi_1$ or $\varphi_2$), then it is capable of acquiring the which-way information of every individual atom by such an operation purely working upon the atom itself, without the trouble to count the photon number (0 or 1) in the micromaser cavity. This, of course will bring more flexibility in designing good measurement schemes to test the wave-particle duality with high fidelity and precision.

In principle, almost all particles, including massless particles such as photons, and massive particles such as electrons, protons, neutrons, atoms, and molecules, all have intrinsic internal freedoms. For instance, photons have polarizations, electrons have spins, protons and neutrons also have nucleus spins, while atoms and molecules have many subtle electronic states. Various interferometers constructed and operated based on the above-mentioned schemes can be employed to make a wave-particle duality test experimentally rather than just gedankenly by using state-of-the-art technologies. Subtle differences between massless particles (known as the medium for interaction force) and massive particles (known as the origin of interaction force) might be revealed by these high fidelity and precision wave-particle duality experimental test. These undoubtedly will open a new door to explore the nature of various elementary particles and their composite objects in quantum world more than 100 years after its door was first uncovered by the pioneers of quantum theory, including Einstein, Bohr, Heisenberg, Schrödinger, Dirac, and many others.

## 8. Conclusions and Perspectives

We have made a systematical analysis on various Young's two-slit interferometers and Mach-Zehnder two-arm interferometers for testing wave-particle duality and principle of complementarity by adopting only elementary methodology of quantum mechanics rather than purely conceptual argument and justification. In particular, we focus on the key issues of how to make good measurement of the which-way information (particle nature) and the interference pattern (wave nature) of quantum objects in these interferometers. We have realized that the internal freedoms of quantum object could be used to construct both which-way detector and interference-pattern detector for the center-of-mass motion of atoms. However, the way of arrangement and cooperation of these two sets of measurement instrument in the interferometer would lead to completely different results. We have found that arrangement of good and bad measurement on either property (wave or particle), or inappropriate arrangement of good measurement on both properties (wave and particle) would lead to failure of simultaneous observation of wave and particle behavior in numerous gedanken and realistic experiments in history and uphold the orthodox conceptual knowledge upon the wave-particle duality and the principle of complementarity. Based on such a new angle of view, we are able to design several schemes that allow for simultaneously acquiring the which-way information for an individual quantum object and the high-contrast interference pattern for an ensemble of these quantum objects by placing two sets of measurement instrument that are well separated in space and whose perturbation on each other is negligibly small within a single Young's interferometer or Mach-Zehnder interferometer.

Our new designs of interferometer are drastically different from many interferometers in history and current times in the concept and motivation. As the focus in placed on how to handle the issue of how to make simultaneous good measurement of the wave (via interference pattern) and particle (via which-way information) nature of quantum objects, rather than on the philosophical and conceptual argument of how to observe the, our schemes offer a more practical means to go beyond the complementarity, namely, a way to simultaneously observe the wave and particle behavior of quantum objects. In addition, our schemes also allow for continuous approach to the ideal realm of perfect wave-and-particle observation by adopting state-of-the-art weak measurement technologies in quantum optics and atomic physics. In this regards, the key issue is no longer the technology for making things finer, but rather the courage for bringing change and revolution of concept.

That quantum objects can behave beyond the wave-particle duality and complementarity principle would stimulate new conceptual examination and exploration of quantum theory at a deeper level 100 years after its foundation. Of course, such an important consequence in fundamental physics involved in our theoretical analysis should be subject to highest-level strict experimental test based on advanced quantum optics and atomic physics technologies. Nonetheless, even if these experimental tests would bring new understanding on the wave-particle duality and complementarity principle in quantum world, other more subtle issues such as whether Heisenberg's uncertainty principle can be violated in Einstein-Podolsky-Rosen type experiment [5], whether localism and realism strictly hold in quantum world, whether one can observe simultaneously the wave and particle nature of a single quantum object, and many others, still largely remain mysterious and will continue to puzzle human beings. Yet, the step of human being to explore unknown frontiers and landscapes in science and nature will never come to a stop, and the new understanding of wave-particle duality and complementarity principle surely will shed new light and bring new insights for answering these fundamental questions in quantum world.

**Acknowledgment**

The author acknowledges the long-time financial support from the National Natural Science Foundation of China, Ministry of Science and Technology of China and Chinese Academy of Sciences ever since the first idea of the present work was raised in 1999.

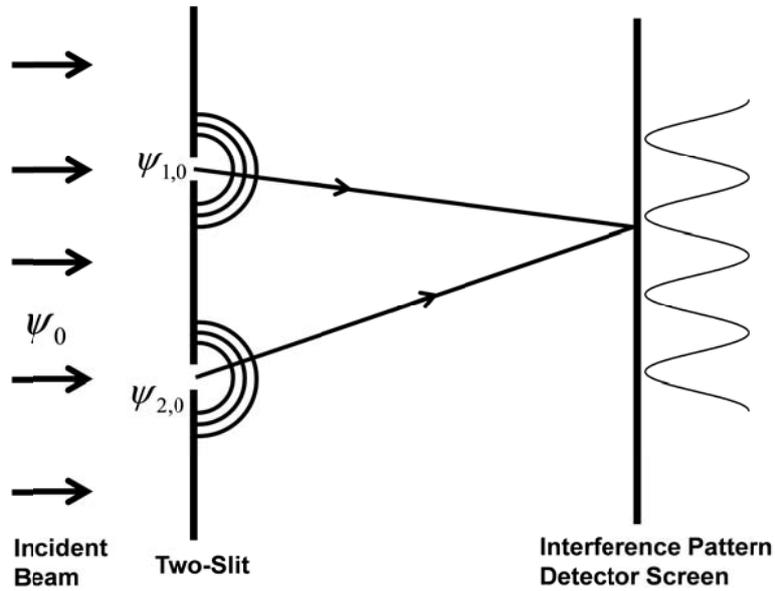

**Fig. 1.** Principle of Young's two-slit interferometer for wave-particle duality test of quantum objects. The interferometer consists of three components: the incident beam of quantum object assumed to have high coherence and described by wave function $\psi_0$; an opaque screen encoded with two parallel slits separated in some distance; a screen directly decorated with detector or through other indirect mechanisms to observe and measure the interference pattern formed by the secondary wave ($\psi_{1,0}$ and $\psi_{2,0}$) diffracting from the two slits. When the slits are fixed, high-contrast interference pattern can be observed while the which-way information, i.e., which slit the quantum objects pass, is unknown completely. When the slits can recoil freely to allow for recording the which-way information, the secondary waves are strongly disturbed so that the high-contrast interference pattern disappears.

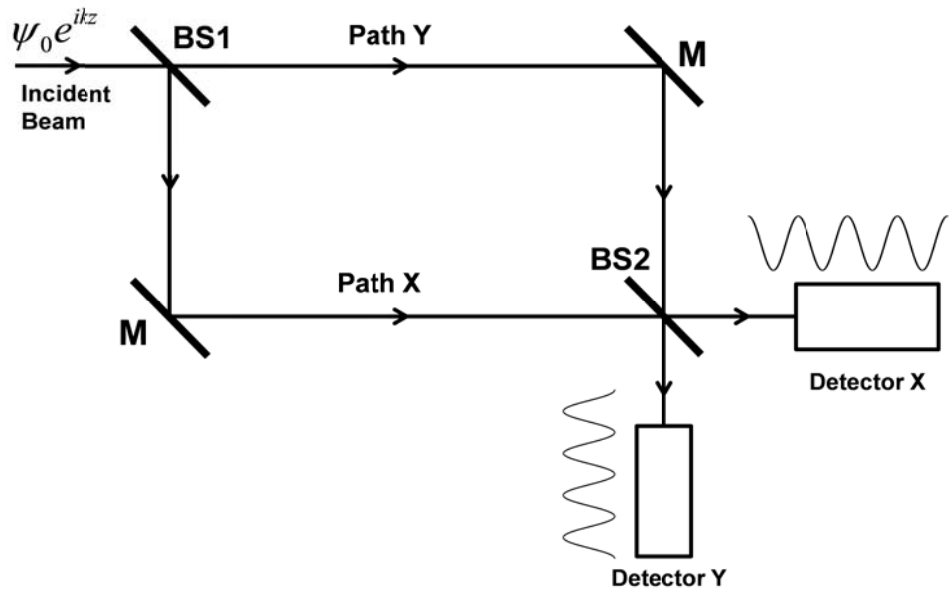

**Fig. 2.** Principle of Mach-Zehnder two-arm interferometer for wave-particle duality test of quantum objects. The incident wave beam of quantum object described passes through the first beam splitter (BS 1) and is separated into two beams, which transport along two paths (path $x$ and path $y$). When they pass through the second beam splitter (BS 2), both detector $x$ and detector $y$ can record the periodic variation of the signal intensity when the path length difference of the two arms is changed continuously, remarking the wave nature of the quantum object. When BS 2 is absent, both detectors observe no periodic variation of signal, but rather a constant value of signal, remarking the particle nature of the quantum object.

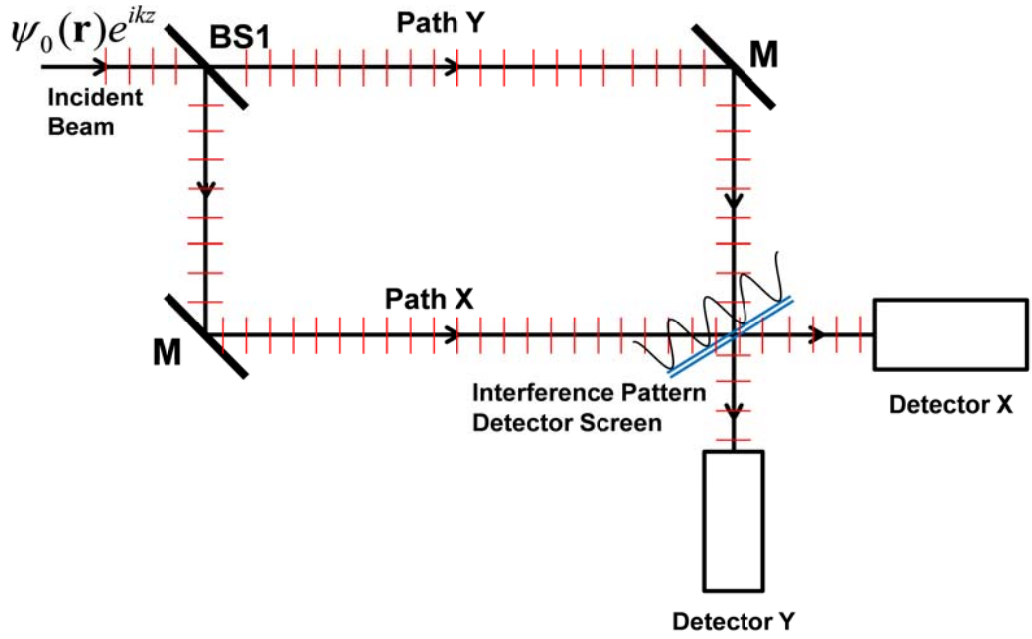

**Fig. 3.** New design of Mach-Zehnder interferometer allowing for simultaneous acquisition of which-way information and high-contrast interference fringe of quantum objects. The incident quantum object wave beam described by $\psi_0(\mathbf{r})e^{i\varphi(z)}$ with longitudinal phase $\varphi(z)$ and transverse wave front $\psi_0(\mathbf{r})=\psi_0(x,y)$ (denoted by the red short line segments) will passes through the first beam splitter (BS 1) and is separated into two beams, which transport along two paths (path $x$ and path $y$). When they intersect in space, interference pattern of wave will appear in the intersection region due to the overlapping of their transverse wave front. A highly transparent (99% transmittance) and weakly scattering or absorption (1% scattering or absorption) screen (denoted by the blue thick lines, instead of BS 2 in Fig. 2) inserted into this region can be used to record the interference pattern of the two wave beams, while imposing only little influence on the propagation of the quantum object. The two wave beam of quantum object will transmit through the screen along their original path and be detected by particle detectors $x$ and $y$, yielding the which-path information of the quantum object. The simultaneous observation of wave and particle nature of quantum objects is made in different space-time domain without apparent mutual repulsion effect seen in conventional Mach-Zehnder interferometer illustrated in Fig. 2.

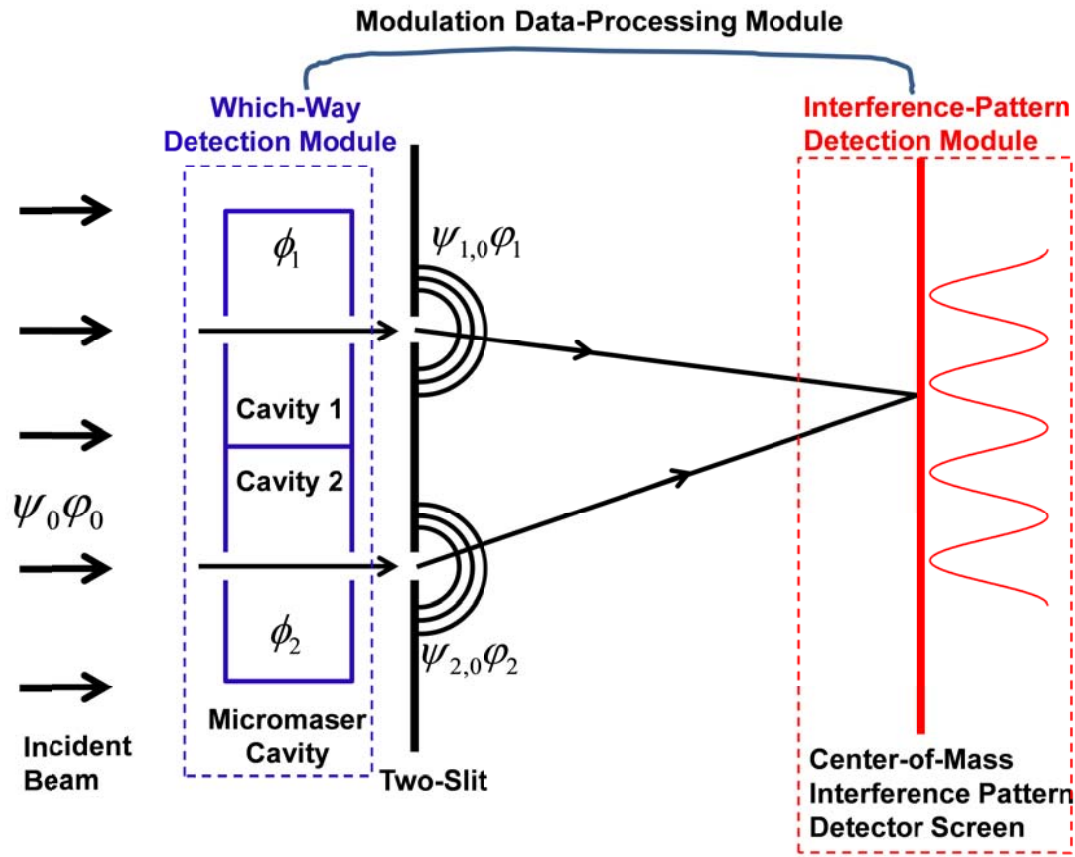

**Fig. 4.** Analysis of Young's two-slit interferometer setup for atoms with internal freedom coupled with the micromaser which-way detectors. The interferometer consists of four components. The first part is the incident Rydberg atomic beam described by composite wave function $\psi_0 \varphi_0$, where $\psi_0$ is the atom center-of-mass wave function assumed to have high coherence and $\varphi_0 = \varphi_e$ is the initial excited quantum state of atomic internal freedom same for all atoms. The second part is the micromaser cavity placed before and collimated with the two slits. They are coupled with the excited quantum state of atom and serve as the which-way detector from their photon quantum state described by $\phi$. By looking at the emitted microwave photon number in each cavity and the corresponding quantum state of field ($\phi_1$ and $\phi_2$) one can tell which cavity the atom passes through. The third part is an opaque screen encoded with two parallel slits separated in some distance and the secondary atom waves at the exit of the two slits are described by $\psi_{1,0}\varphi_1$ and $\psi_{2,0}\varphi_2$, where

$\varphi_1 = \varphi_2 = \varphi_g$ is the final state of atom internal freedom (ground state), while $\psi_{1,0}$ and $\psi_{2,0}$ are the atom center-of-mass wave function. The fourth part is a screen decorated with detectors that allow for only observing and measuring the interference pattern reflecting the atom center-of-mass positioning probability without the complexity induced by incorporation and interaction with the internal freedom of atom. In practical experiment, the second part and fourth part can be categorized as the which-way detector module (blue box) and the interference-pattern detection module (red box), respectively. These two modules act together in well-separated space-time domain without mutual influence and can allow one to simultaneously acquire the wave and particle behavior of the atoms. However, if another module called the modulation data-processing module is used to modulate artificially the atom center-of-mass interference pattern with the coincidence-counting outcome of cavity photon numbers in the which-way detection module, the outcome of this joint data-convolution action will lead to seemingly apparent disappearance of interference pattern because it just corresponds to the mathematical interference term of $\text{Re}[(\varphi_1^* \varphi_2)(\phi_1^* \phi_2) \psi_1^*(\mathbf{R}) \psi_2(\mathbf{R})] = 0$. This result has been popularly described as the entanglement enforced complementarity principle by Scully and many others. Besides, such an artificial connection of the which-way detection module, where many changes and modifications can be made, with the interference-pattern detection module is also the origin of so-called delayed-choice, quantum eraser, and many other counter-intuition conceptual arguments and deductions.